\documentclass{aa}  

\usepackage{graphicx}
\usepackage{txfonts}
\begin{document}

   \title{Data-driven dissection of emission-line regions in Seyfert galaxies}


   \author{Beatriz Villarroel
          \inst{1}
          \and
          Andreas J. Korn\inst{1}
          }

  \institute{Department of Physics and Astronomy, Uppsala University, \\
              SE-751 20 Uppsala, Sweden\\
              \email{beatriz.villarroel@physics.uu.se, andreas.korn@physics.uu.se}
       }

   \date{Received ; accepted }

 
  \abstract
   {}
   {Indirectly resolving the line-emitting gas regions in distant Active Galactic Nuclei (AGN) requires both high-resolution photometry and spectroscopy (i.e. through reverberation mapping). Emission in AGN originates on widely different scales; the broad-line region (BLR) has a typical radius less than a few parsec, the narrow-line region (NLR) extends out to hundreds of parsecs. But emission also appears on large scales from heated nebulae in the host galaxies (tenths of kpc).}
   {We propose a novel, data-driven method based on correlations between emission-line fluxes to identify which of the emission lines are produced in the same kind of emission-line regions. We test the method on Seyfert galaxies from the Sloan Digital Sky Survey (SDSS) Data Release 7 (DR7) and Galaxy Zoo project.}
   {We demonstrate the usefulness of the method on Seyfert-1s and Seyfert-2 objects, showing similar narrow-line regions (NLRs). Preliminary results from comparing Seyfert-2s in spiral and elliptical galaxy hosts suggest that the presence of particular emission lines in the NLR depends both on host morphology and eventual radio-loudness. Finally, we explore an apparent linear relation between the final correlation coefficient obtained from the method and time lags as measured in reverberation mapping for Zw229-015.}
   {}


   \maketitle
%

\section{Introduction}\label{sec:intro}

Quasars, radio galaxies, Seyfert-1 and Seyfert-2 objects, low-ionization nuclear emission-line region (LINERs) and blazars all belong to the rich zoo of active galactic nuclei (AGN). One of the most central questions in astrophysics concerns which physical mechanisms dominate the emission-line production in the various AGN classes. 
Radio galaxies are explained in terms of jet physics (Baum et al. \citeyear{Baum1995}); LINERs are commonly suspected to be dominated by shocks (Dopita \& Sutherland \citeyear{Dopita1995}) 

AGN Unification theory (Antonucci \citeyear{Antonucci1993}) predicts the same underlying physical mechanisms in the center of Seyfert-1 and Seyfert-2 nuclei. The strong ultraviolet light from the photoionizing central engine heats the surrounding gas and is absorbed by the dusty torus, processed and reemitted in the infrared. In the case of Seyfert-1s one sees into the central engine without the torus obscuring the view. In Seyfert-2s, the view is believed to be obscured; one sees the line-emitting region outside the dusty torus, giving rise to narrow Balmer emission and strong forbidden lines so typical of AGN spectra. This region is commonly referred to as the ``narrow-line region'' (NLR).

There are several open questions related to the physics of the NLRs. The first is the most basic: is it the same physics causing the narrow-line emission in all Seyfert galaxies? The presence of hidden broad-line regions in many Seyfert-2s supports a common central engine and that the answer is a resounding `yes'. But again one may ask: is the narrow-line emission isotropically distributed around the central engine? And is the narrow-line region independent of the physics of the host galaxy? Or could the host galaxy play an important role too?

The obstacles of disentangling the liable mechanisms bring a challenge upon astronomers often dealt with through spectral decomposition or theoretical modeling of spectral energy distributions (SED). The continuum source itself, the AGN corona, the broad-line region, the narrow-line region outside the dusty torus, and finally the heated gas in the host galaxy itself, all contribute to the overall galaxy spectra. 

We herein propose a novel method to study emission-line regions in statistical
samples of galaxies. Except for the assumption about interdependence of emission lines, the method uses no other physical assumptions. The basis of the method are correlation coefficients between emission lines, described in Section 2.1. We demonstrate the method on spectra of Seyfert-1 and Seyfert-2 galaxies. We test if the method can provide a new way of estimating distances between the line-emitting region and the black hole in AGN. In Section 3, we summarize our results in Section 4.

\section{Methods}
\subsection{Co-locative Correlation Analysis}

While a correlation between two emission line fluxes in a sample 
does not imply causation, we $assume$ that if fluxes from two emission lines correlate the same way
with all the other observed line fluxes, the two emission lines formed 
due to the same underlying physical mechanism. 

As a more explicit example, let's say, in a sample of 1000 galaxies we observe 
an emission line $A1$ and see that its flux strongly correlates with the flux of 
emission lines $A3$, $A5$, $A6$ and $A7$. The line also anticorrelates with emission line $A10$. It does not correlate with any other lines. Incidentally, we observe 
another emission line $A2$ and find out that it correlates in a similar fashion (as $A1$) with emission lines $A3$, $A5$, $A6$ and $A7$. As in the case of $A1$, it also shows anticorrelation with $A10$, and no correlations with other lines. The example is illustrated in the matrix below, showing values $c$ of correlation coefficients. Thus, $c$ ranges from 0 to 1. If including only strong correlations, $c$ ranges from 0.7 to 1 in value.

\[ \begin{array}{ccccccccccccc}
   & & A1 & A2 & A3 & A4 & A5 & A6 & A7 & A8 & A9 & A10\\
l_{1} & A1 & 1 & c & c & 0 & c & c & c & 0 & 0 & -c\\
l_{2} & A2 & c & 1 & c & 0 & c & c & c & 0 & 0 & -c\\
l_{3} & A3 & c & c & 1& & & & & & &\\ 
l_{4} & A4 & 0 & 0 & & 1 & & & & & &\\
l_{5} & A5 & c & c & & & 1 & & & & &\\
l_{6} & A6 & c & c & & & & 1& & & &\\
l_{7} & A7 & c & c & & & & & 1& & &\\
l_{8} & A8 & 0 & 0 & & & & & & 1 & &\\
l_{9} & A9 & 0 & 0 & & & & & & & 1&\\
l_{10} & A10 & -c & -c & & & & & & & &\\ \end{array}\]

Given the similarity in behaviour, the likelihood that $A1$ and $A2$ would form due to 
different underlying physics becomes vanishingly small. $A1$ and $A2$ have likely formed in
the same gas regions (or physical substructure).

Of course some emission lines do not appear above/below a certain critical temperature or density; not all emission lines from the same region will be interdependent on each other. But those emission lines that are, as $A1$ and $A2$ in the example, are assumed to have a common origin.

We develop this idea into an algorithm calculating the correlations-of-correlations between emission-line fluxes in spectra using Spearman coefficients, 
hereinafter referred to as Co-locative Correlation Analysis (CoCoA). In practice this is done by calculating the correlation coefficient between
rows $l_{i}$ in the matrix. As an example, calculating the correlation-of-correlations between $A1$ and $A2$ is done by estimating the 
correlation coefficient of $l_{1}$ and $l_{2}$. In CoCoA, we always specify a line of interest as a reference e.g. [\ion{O}{iii}]5007 or H$\alpha$.
These two lines are suitable reference lines as they are typically strong in Seyfert galaxies.

To reduce noise and detect only the very strongest correlations based on Spearman coefficients, we set $|\rho|$ $>$ 0.7, so that correlations with $|\rho|$ $<$ $c_{lim}$ or the correlation-of-correlation coefficient $|\rho'|$ $<$ $c'_{lim}$ are set equal to zero, and where $c_{lim}$ and $c'_{lim}$= 0.7 The accompanying standard errors $\epsilon_{\rho}$ (error of correlation) and $\epsilon_{\rho'}$ (error of correlation-of-correlations) are calculated via bootstrapping resampling and we set $|\rho|$ $>$ 3$\epsilon_{\rho}$ to exclude weak or insignificant correlations for estimating $|\rho'|$. For the final step, we also require $|\rho'|$ $>$ 1$\epsilon_{\rho'}$ to exclude poor estimates of the correlation-of-correlations.

\subsection{Galaxy samples}

To investigate the usefulness of CoCoA, we turn to observations. Galaxy spectral line info are taken from the Sloan Digital Sky Survey (SDSS) (York et al. \citeyear{York2000}) Data Release 7 (Abazajian et al. \citeyear{Abazajian2009}).

\subsubsection{Parent samples}

We select objects within spectroscopic redshift 0.03 $< z <$ 0.2 and require minimum SDSS Gaussian line heights $h$(H$\alpha$) $>$ 10 * 10$^{-17} $erg/s/cm$^{2}/\AA$ and $h$(H$\beta$) $>$ 5 * 10$^{-17} $erg/s/cm$^{2}/\AA$ to minimize contamination effects of stellar absorption. Using Gaussian H$\alpha$ line width and optical emission line diagnostics, we classify the selected AGN into ``Type-1'' if they have $\sigma$(H$\alpha$) $>$ 10 \AA\  . ``Type-2'' AGN have $\sigma$(H$\alpha$) $<$ 10 \AA\  and fulfil the criterion (Kauffmann et al. 
\citeyear{Kauffmann2003}):

\begin{equation}
\log([\ion{O}{iii}]/{\rm H}\beta) > 0.61/(\log([\ion{N}{ii}]/{\rm H}\alpha))-0.05)+1.3 \label{eq,BPT}
\end{equation}\label{BPTequation}

We remove all objects fulfilling (Kewley et al. \citeyear{Kewley}):

\begin{equation}
\log([\ion{O}{iii}]/{\rm H}\beta) < 0.61/(\log([\ion{N}{ii}]/{\rm H}\alpha))-0.47)+1.19 \label{eq,LINER}
\end{equation}\label{LINER}

In these AGN samples LINERs and strong Seyferts are removed but the samples are biased towards composite objects. However, they are suitable for probing the NLR with CoCoA. The interest in using weaker AGN is due to concerns of NLR anisotropies at higher luminosities when the Unification might break. LINERs, on the other hand, might be dominated by shock physics and complicate the analysis if mixed in. It is difficult to foresee how this could impact CoCoA, the chosen sampling likely minimizes related issues.

From the AGN, we select two samples with the help of Galaxy Zoo 1 \& 2 (Lintott et al. \citeyear{Lintott,Lintott2010}, Willett \citeyear{Willett}) that are face-on spiral hosts and two samples of elliptical hosts:

\begin{enumerate}

\item 148 face-on spiral Type-1 AGN
\item 3488 face-on spiral Type-2 AGN
\item 40 elliptical Type-1 AGN
\item 168 ``ordinary'' elliptical Type-2 AGN

\end{enumerate}

Matching the face-on spiral-host Type-1 and Type-2 AGN by redshift $z$ and $L$[\ion{O}{iii}] and calculating the average Gaussian line width for the samples one can see they have negligible difference in [\ion{O}{iii}] line width between Type-1s ($\log_{10}$($\sigma$)= 0.3945 $\pm$ 0.028) and Type-2s ($\log_{10}$($\sigma$)= 0.4042 $\pm$ 0.0191). This difference is smaller than the line width difference reported ($\sim$ 1.8 $\sigma$) in the samples of (Villarroel et al. \citeyear{VillarroelNyholm2015}) where a small, but potential anisotropy in $\sigma$[\ion{O}{iii}]5007 is present. Therefore, the samples in this study have a smaller risk of anisotropies in NLR kinematics. The samples are used primarily to test the possibilities and limitations of CoCoA. It is therefore important to note that no conclusions regarding the Unification theory can be obtained from this study. Hereafter, we always use and refer to the parent samples.

In Section 3.1 we use the face-on spiral Type-1 and Type-2 AGN samples (samples 1 \& 2) to compare the NLRs of Seyfert-1 and Seyfert-2 AGN.

In Section 3.2 we compare the NLRs of Seyfert-2 AGN in spiral hosts vs Seyfert-2 AGN in elliptical hosts (samples 2 \& 4). All four parent samples are used for a first test of the method with results shown in Table 1 (Section 3).

\subsubsection{Narrow-line radio galaxies}

From the Faint Images of the Radio Sky at Twenty cm (FIRST) catalogue, we select sources via the SDSS Casjobs interface with an integrated radio flux larger than 5 mJy and spectroscopic redshift 0.03 $< z <$ 0.2. From these we keep only elliptical-host objects with narrow $\sigma($H$\alpha) < $ 10 \AA , leaving 358 objects.

In Section 3.2 we use the narrow-line radio galaxies (NLRGs) in elliptical hosts to compare with the ``ordinary'' (presumably radio-quiet) elliptical-host Type-2 AGN selected in Section 2.2.1.

\subsubsection{Matching of samples}

In Section 3.1, the Seyfert-1s and Seyfert-2s are pairwise matched in the following properties, as in 
Villarroel et al. \citeyear{VillarroelNyholm2015}: (a) redshift $z$, (b) $z$ and Balmer decrement H$\alpha$/H$\beta$ and (c) $z$ and [\ion{O}{iii}]5007, leading to indistinguishable redshift distributions between Type-1 and Type-2 AGN and identical sample sizes. A Kolmogoroff-Smirnoff test is performed for all matched properties to ensure similarity between the underlying distributions.

In Section 3.2. we match the samples in Balmer decrement $F$(H$\alpha$/H$\beta$) and redshift. First, we match and compare spiral to elliptical hosts Type-2 AGN (sample sizes $N$ = 163). Then we match and compare ``ordinary'' elliptical to radio-loud elliptical Type-2 AGN (sample sizes $N$ = 28).

\subsection{Defining CoCoA group 1 \& 2}

For simplicity, we define ``CoCoA group 1'' as the emission lines with $|\rho'|>$ 0.7 calculated using H$\alpha$ as a reference point in Type-1 objects, representing the broad-line region. The ``CoCoA group 2'' uses [\ion{O}{iii}]5007 as a reference point
and instead represents the narrow-line region lines in both Type-1 and Type-2 objects.

An important note: if an emission line is absent from a particular group of lines it does $not$ necessarily mean the line emission does not form simultaneously with others in the same physical process. But it $does$ mean a larger fraction of the observed line emission is formed elsewhere. Thus, CoCoA primarily picks the least contaminated lines from selected emission-line regions. A list of all SDSS lines used in our analysis can be found in Table \ref{JustLines}. It consists of all lines for which measurements in SDSS DR7 existed for the AGN.

\begin{table*}[ht]
\caption{Lines used in the CoCoA-analysis.}
\centering
\begin{tabular}{c}
\multicolumn{1}{c}{Absorption/emission line} \\
\hline
\hline
{[}\ion{Ne}{v}] 3347\\
{[}\ion{Ne}{v}] 3427\\
{[}\ion{O}{ii}] 3727\\
{[}\ion{O}{ii}] 3730\\
\ion{He}{i} 3889\\
{[}\ion{S}{ii}] 4072\\
H$\gamma$ 4342\\
H$\delta$ 4103\\
{[}\ion{O}{iii}] 4364\\
H$\beta$ 4863\\
{[}\ion{O}{iii}] 4960\\
{[}\ion{O}{iii}] 5007\\
H$\alpha$ 6565\\
{[}\ion{N}{ii}] 6550\\
{[}\ion{N}{ii}] 6585\\
{[}\ion{O}{i}] 6302\\
{[}\ion{O}{i}] 6366\\
{[}\ion{S}{ii}] 6718\\
{[}\ion{S}{ii}] 6733\\
\ion{Ca}{ii} 8500\\
\ion{Ca}{ii} 8544\\
\ion{Ca}{ii} 8665\\[0.2ex]
\hline
\end{tabular}
\label{JustLines}
\end{table*}

\section{Results}

We run CoCoA on the parent samples. We calculate the correlations-of-correlations $\rho'$ of all emission-lines with respect to particularly [\ion{O}{iii}]5007, believed to originate in the NLR. The results are collected in Table \ref{NLRcombined}. At first glance the differences between the samples in the clustering of lines with [\ion{O}{iii}]5007 are striking: all four AGN classes show different emission lines in their NLR. The Seyfert-1 and Seyfert-2 appear different, e.g. spiral Seyfert-1 have [\ion{O}{i}]6302 while spiral Seyfert-2 do not. But selection criteria are critical and the samples differ in size. This will strongly influence the first step when simple correlations coefficients are calculated. Also, Balmer lines are expected to be dominated by BLR emission in Seyfert-1s.

\begin{table*}[ht]
\caption{Parent samples. CoCoA strengths $|\rho'|$ of CoCoA group 2 in the parent samples. Reference is [\ion{O}{iii}]5007. All lines with $|\rho'|<$0.7 are disregarded.
The errors indicated are estimates of the standard deviation to $|\rho'|$ from bootstrap resampling.}
\centering
\begin{tabular}{c c c c c}
\multicolumn{5}{c}{Type-1 Samples, NLR} \\
\hline\hline
Emission-line & Spiral Type-1 (148) & Elliptical Type-1 (40) & Spiral Type-2 (3488) & Elliptical Type-2 (168)\\[0.2ex]
\hline
H$\alpha$ & ... & ... & 0.8803 $\pm$ 0.0552 & ... \\
H$\beta$ & ... & ... & 0.9053 $\pm$ 0.0493 & 0.8618 $\pm$ 0.0794 \\
H$\gamma$ & -& ... & ... & ... \\
H$\delta$ & -& ... & ... & ... \\
{[}\ion{O}{i}] 6302 & 0.8746 $\pm$ 0.0520 & ... & ... & 0.7710 $\pm$ 0.0875\\
{[}\ion{O}{i}] 6366 & ... & ... & ... & ... \\
{[}\ion{O}{ii}] 3727 & ... & ... & ... & ... \\
{[}\ion{O}{ii}] 3730 & ... & ... & ... & ... \\
{[}\ion{O}{iii}] 4364 & ... & ... & ... & ... \\
{[}\ion{O}{iii}] 4960 & 0.9690 $\pm$ 0.0278 & 0.9473 $\pm$ 0.0497  & ... & ... \\
{[}\ion{N}{ii}] 6550 & ... & ... & 0.8701 $\pm$ 0.0596 & ... \\
{[}\ion{N}{ii}] 6585 & 0.7667 $\pm$ 0.1297 & 0.7320 $\pm$ 0.1659 & 0.8724 $\pm$ 0.0582 & ... \\
{[}\ion{S}{ii}] 4072 & ... & ... & ... & ... \\
{[}\ion{S}{ii}] 6718 & 0.9588 $\pm$ 0.0415 & 0.7019 $\pm$ 0.1511 & 0.8876 $\pm$ 0.0557 & 0.8271 $\pm$ 0.0837\\
{[}\ion{S}{ii}] 6733 & 0.9644 $\pm$ 0.0350 & 0.8210 $\pm$ 0.0963 & 0.8914 $\pm$ 0.0544 & 0.7944 $\pm$ 0.0992\\[0.2ex]
\hline\hline
\multicolumn{3}{c}{Other lines} \\
\hline
{[}\ion{Ne}{v}] 3347 & ... & ... & ... & ... \\
{[}\ion{Ne}{v}] 3427 & ... & ... & ... & ... \\
\ion{He}{i} 3889 & ... & ... & ... & ... \\
\ion{Ca}{ii} 8500 & -0.8156 $\pm$ 0.0998 & ... & ... & ... \\
\ion{Ca}{ii} 8544 & ... & ... & -0.8334 $\pm$ 0.0548 & ... \\
\ion{Ca}{ii} 8665 & ... & ... & -0.8176 $\pm$ 0.0561 & ... \\
\end{tabular}
\label{NLRcombined}
\end{table*}

\subsection{The narrow-line region in Seyfert-1s and Seyfert-2s}

Seyfert-1 and Seyfert-2 galaxies are believed to have the same photoionizing central engine, 
but are observed from different viewing angles relative to the central source (Antonucci \citeyear{Antonucci1993}).
But some statistical studies find differences in clustering of neighbours around the two types of AGN (Dultzin et al. \citeyear{Dultzin}, Koulouridis et al. \citeyear{Koulouridis2013}, Jiang et al. \citeyear{Jiang2016}) and type of neighbours (Villarroel \& Korn \citeyear{VillarroelKorn2014}). Other models propose Seyfert-2s might have significantly more star formation (Maiolino et al. \citeyear{Maiolino1995}), supported recently by Villarroel et al. (\citeyear{VillarroelNyholm2015}) who find Seyfert-2 hosts have younger stellar populations. It opens up a remote possibility for different central engines in Seyfert-1 and Seyfert-2 galaxies where nuclear star formation strongly contributes to the narrow-line region in Seyfert-2s.

If the central engine is the same, the same physical mechanisms cause the narrow-line emission. And if so, we expect CoCoA to show the same lines falling into the narrow-line region. We wonder whether in any of the matched samples, the supposed narrow-line region of Seyfert-1 and Seyfert-2 galaxies include the same set of lines. The Balmer lines will be disregarded in the comparison as they are dominated by different emission-line regions in Seyfert-1 (dominated by BLR lines) and Seyfert-2 spectra (dominated by NLR lines). Some other problematic lines are the \ion{Ca}{ii} absorption lines and the [\ion{Ne}{v}]3347,3427, as they despite being used in the correlations-of-correlations calculations in $\sim$ 2/3 of the cases lack measurements. We will neglect them when comparing $\rho'$ values of Seyfert-1s and Seyfert-2s.

\begin{figure*}
 \centering
   \includegraphics[scale=.8]{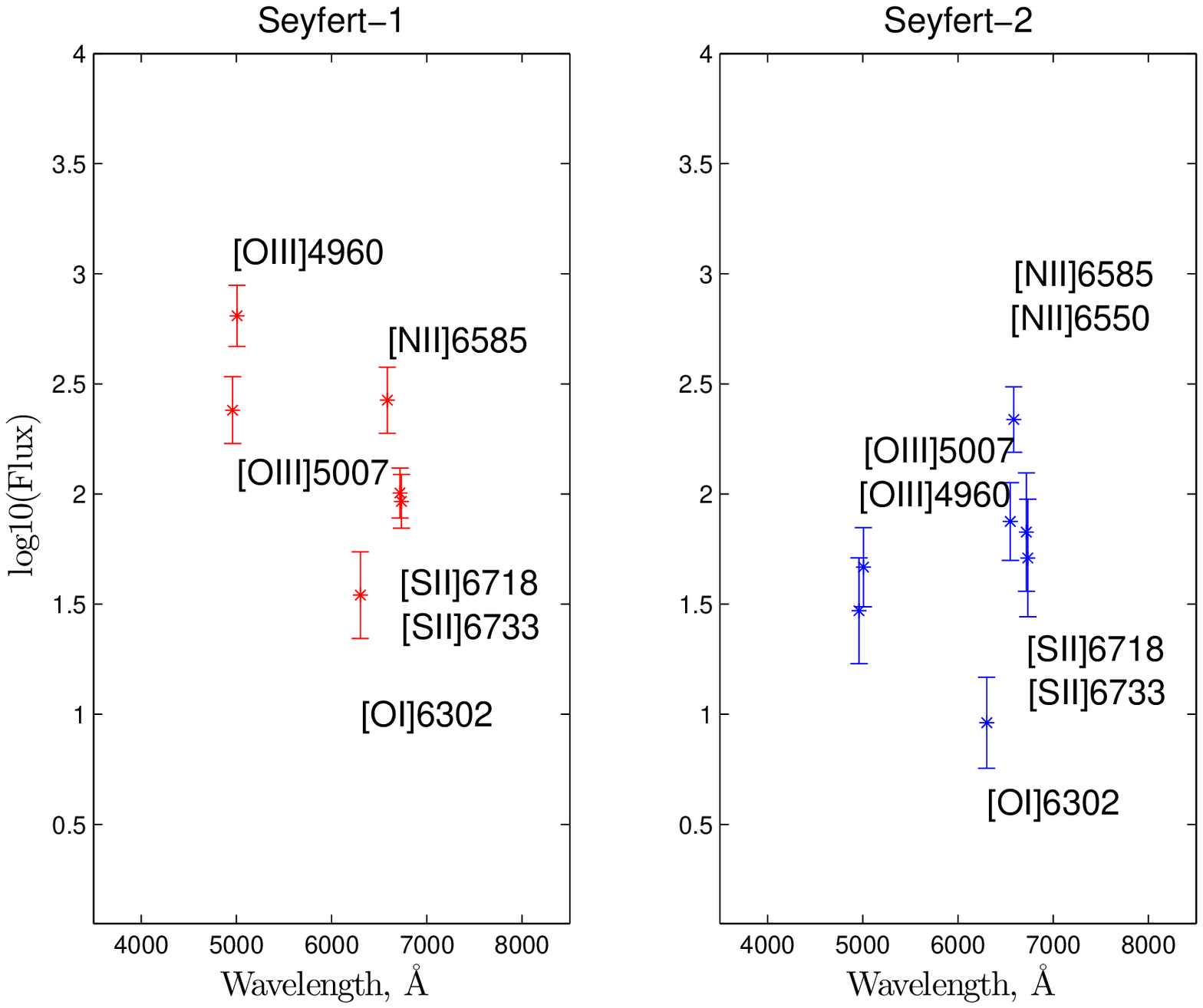}
     \caption{The fluxes of emission lines in the CoCoA group 2. The left graph shows CoCoA group 2 fluxes with 3$\sigma$ error bars for 87 Seyfert-1s.
The right graph shows CoCoA group 2 fluxes for 87 Seyfert-2s, selected to match in redshift and Balmer decrement H$\alpha$/H$\beta$ to the Seyfert-1s, Balmer lines excluded. The CoCoA group 2 consists of the same lines in both cases. Seyfert-1s appear however one order of magnitude more luminous in the [\ion{O}{iii}] lines and [\ion{O}{i}]6302 than their Seyfert-2 counterparts. Note as the broadening of the H$\alpha$-line takes place in Seyfert-1s, the [\ion{N}{ii}]6550 line becomes unresolved. The other matchings do not produce as good agreement in the CoCoA groups 2. In case of $L$[\ion{O}{iii}]5007-matched samples the CoCoA group 2 regions in Seyfert-1 and Seyfert-2s show very different line composition from each other, suggesting the matching is unphysical. The fluxes are in units of 10$^{-17}$ erg/s/cm$^{2}$.}
               \label{SeyfertCoCoa2}
     \end{figure*}

We run CoCoA to compare the CoCoA group 2 (ignoring the Balmer lines). Samples matched in only $z$ or in $z$ and [\ion{O}{iii}]5007 give different lines in CoCoA group 2. But samples matched in dust extinction and redshift produce the same composition of the CoCoA group 2 with identical lines as expected for the narrow-line regions in the most general Unification theory, see Figure \ref{SeyfertCoCoa2}. It shows the cospatial existence of [\ion{O}{iii}]5007 with lines such as [\ion{S}{ii}]6718,6733, [\ion{N}{ii}]6550,6558 and [\ion{O}{i}]6302. But as found in Villarroel et al. (submitted), the difference in $L$[\ion{O}{iii}]5007 is strong and Seyfert-1s are much more luminous than their Seyfert-2 counterpart. One may wonder if the effect is not caused by an erroneous decomposition of the [\ion{O}{iii}] emission lines due to the underlying broad H$\beta$. An erroneous decomposition could lead to differences in the flux-normalized average line width $\sigma$[\ion{O}{iii}]5007 in $L$[\ion{O}{iii}]5007-matched samples. However, the flux-normalized average line width $\sigma$[\ion{O}{iii}]5007 is the same for Seyfert-1s and Seyfert-2s, suggesting the observed effect is not caused by any problems in decomposition.

Interestingly, the CoCoA group 2 is characterized by the absence of the $L$[\ion{O}{ii}]3727,3730 lines in both types of AGN, suggesting the $L$[\ion{O}{ii}] in Seyferts is dominated by processes related to star formation.

However, comparing two classes of objects to each other with CoCoA reveals that small sampling effects and problems with
lines that only sometimes appear e.g. [\ion{Ne}{v}]3347,3427 and \ion{Ca}{ii} lines make it difficult to draw firm conclusions based
only on a single CoCoA run. A particular selection sometimes results in the presence of one line, that disappears in a different selection. 
Perhaps a more complete picture can be achieved when directly comparing two object classes if 
running a set of physically motivated samplings, only considering the lines that appear in CoCoA group 2 in all samples.
It remains an open problem how to optimize CoCoA so that robust conclusions can be drawn when comparing two groups
of objects.

\subsection{Spiral vs elliptical Seyfert-2s}

Since the physics behind the narrow-line regions of Seyfert-1s and Seyfert-2s appears to be the same,
we ask ourselves the physics of NLR depends on the AGN type or also can be influenced by
the properties of the host galaxy itself. Normally, samples of AGN in elliptical host galaxies are dominated by radio-loud objects, but as our Seyfert-2s are selected by the emission-line ratio diagrams and have a 
lower cut in the $EW$(H$\alpha$) and $EW$(H$\beta$) values, we push them towards being dominated by radio-quiet AGN. The question posed above can be approached by comparing the narrow-line regions in elliptical-host Seyfert-2 AGN to spiral-host Seyfert-2 AGN.

\begin{figure*}
 \centering
   \includegraphics[scale=.8]{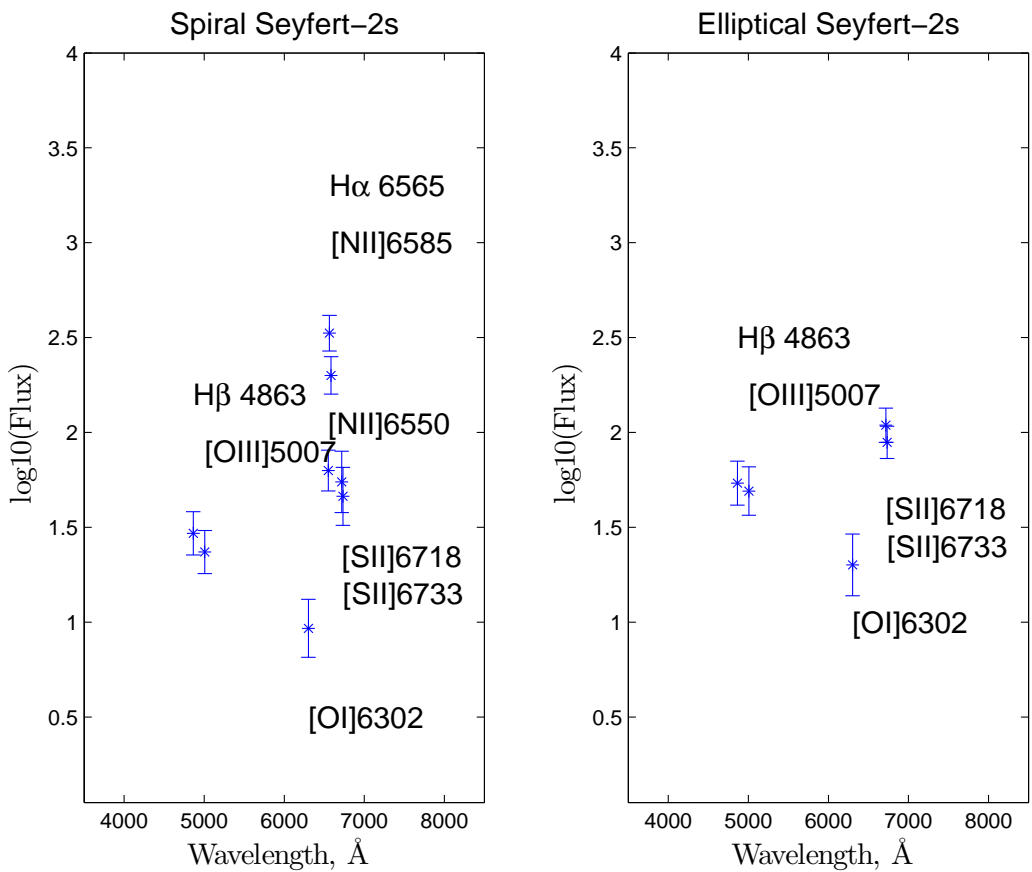}
     \caption{The fluxes of emission-lines in CoCoA group 2. The left graph shows CoCoA group 2 fluxes with 3$\sigma$ error bars for 163 Seyfert-2s in spiral hosts. The right graph shows CoCoA group 2 fluxes for 163 Seyfert-2s in elliptical hosts, 
selected to match the first sample in redshift and Balmer decrement H$\alpha$/H$\beta$. The fluxes are in units of 10$^{-17}$ erg/s/cm$^{2}$.}
               \label{SeyfertMorf}
     \end{figure*}

We run the code on dust extinction-matched Seyfert-2s in spiral and elliptical hosts galaxies, respectively, see Table \ref{NLRmorphology} for results.
H$\alpha, $H$\beta$, [\ion{O}{i}]6302, [\ion{N}{ii}]6550,6586 and [\ion{S}{ii}]6718,6733 are detected in the CoCoA group 2 of spiral-host Seyfert-2 galaxies. 
The elliptical-host Seyfert-2s do not have H$\alpha$ or [\ion{N}{ii}]6550,6586 in their CoCoA group 2. We compare the two sets of NLR fluxes 
in Figure \ref{SeyfertMorf} and see that all fluxes are significantly (size of effect $>$3$\sigma$) stronger in the elliptical hosts compared to the spiral hosts. 
Does this difference in the NLR mean that the nature of the NLR depends on the morphology of host galaxies or could it be
that most of our elliptical hosts are radio-loud AGN in fact? If we compare the elliptical Type-2 AGN to 
NLRGs, their CoCoA groups 2 differ. Thus, the difference between the 
narrow-line regions in spiral-host Type-2 AGN versus elliptical host Type-2 AGN cannot be explained by the 
radio-loud/radio-quiet dichotomy alone. The morphology of the host must matter to some extent for the emission-line properties of the narrow-line region. Perhaps, this means that the properties of the dusty torus -- dust covering factor (Elitzur \citeyear{Elitzur2012}, Ramos-Almeida et al. \citeyear{Cristina}, Ricci et al. \citeyear{Ricci}) or the number of components (Tristram et al. \citeyear{Tristram2014}) -- is connected to the galaxy morphology.

\begin{table*}[ht]
\caption{Elliptical vs spiral Type-2 AGN. The CoCoA strengths $|\rho'|$ for lines in the NLR in dust- and redshift matched samples are tabulated. The reference sample is marked with a star (*). [\ion{O}{iii}]5007 is used as
a reference and the individual CoCoA strengths are listed, using $|\rho'|>$0.7 as lower limit.}
\centering
\begin{tabular}{c c c}
\multicolumn{3}{c}{Samples} \\
\hline\hline
Emission-line & Elliptical* (163) & Spiral (163)\\[0.2ex]
\hline
H$\alpha$ & ... & 0.8756 $\pm$ 0.0629\\
H$\beta$ & 0.7996 $\pm$ 0.175 & 0.9353 $\pm$ 0.0525\\
{[}\ion{O}{i}] 6302 & 0.8358 $\pm$ 0.0823 & 0.7271 $\pm$ 0.1311\\
{[}\ion{N}{ii}] 6550 & ... & 0.8641 $\pm$ 0.0671\\
{[}\ion{N}{ii}] 6585 & ... & 0.8667 $\pm$ 0.0659\\
{[}\ion{S}{ii}] 6718 & 0.8091 $\pm$ 0.1064 & 0.8904 $\pm$ 0.0604\\
{[}\ion{S}{ii}] 6733 & 0.7320 $\pm$ 0.1331 & 0.9046 $\pm$ 0.0552\\[0.2ex]
\hline
\end{tabular}
\label{NLRmorphology}
\end{table*}

A difference in the NLR could perhaps be seen in eventual differences in line ratios. 
The temperature in a nebular cloud might be tricky to measure as the most reliable method 
based on three [\ion{O}{iii}] emission lines, namely the [\ion{O}{iii}]4960, [\ion{O}{iii}]5007 and [\ion{O}{iii}]4364 e.g. Osterbrock (\citeyear{Osterbrock}) while the CoCoA groups 2 do not include the [\ion{O}{iii}]4960 line. The absence of this line can indicate that the [\ion{O}{iii}]4960 is contaminated by emission from other parts of the galaxy and that the flux measurement is not reliable enough to give a useful estimate. But it can also mean that most of the [\ion{O}{iii}]4364 is formed in a higher-density region as in a two-zone model (Baskin \& Laor \citeyear{Baskin2005}).

On the other hand, the rates of collisional de-excitations of two lines from the same ion with almost identical excitation energies depend strongly on electron density $n_{e}$. The [\ion{S}{ii}]6718 and [\ion{S}{ii}]6733 lines are good examples of useful lines and we calculate the line ratios $R$ = $\frac{I([\ion{S}{ii}]6718)}{I([\ion{S}{ii}]6733)}$ giving $R_{spiral}$=1.19 and $R_{elliptical}$=1.23. Associated errors based on taking the standard error from the $\log_{10}$ of the distributions are large and no significant difference in electron density is seen. Assuming a temperature $T$ = 10 000 K, this electron density corresponds to $n_{e}\sim 4 \cdot 10^{2} cm^{-1}$. We note that the slight difference in the $\frac{I([\ion{O}{iii}5007)]}{I(H\beta)}$ probes indirectly the ionization parameter $U$ where the median value is $\frac{I([\ion{O}{iii}5007)]}{I(H\beta)}$=0.66 for spirals and $\frac{I([\ion{O}{iii}5007)]}{I(H\beta)}$=0.75 for ellipticals. Due to the strong asymmetry of the underlying distribution we do not obtain reliable confidence intervals. The low value shows that many of the AGN in our samples have weak central engines.

\subsection{Can one measure time lags with CoCoA?}

Assuming the gas is gravitationally bound to the compact object in an AGN (Gaskell \citeyear{Gaskell1988}),
one can estimate the virialized mass of the central super-massive black hole (SMBH) using the distance $R$ between the line-emitting region and the black hole.

Traditionally, two different approaches for estimating $R$ are the photoionization method for any AGN (Netzer \citeyear{Netzer1990}) or reverberation mapping for variable ones (Gaskell \citeyear{Gaskell1988}, Peterson \citeyear{Peterson1993}). While the photoionization
method suffers from many uncertainties, the reverberation mapping is perhaps more accurate
but also more restricted and costly. 

In reverberation mapping (RM), multi-epoch observations of variable AGN permit to measure the time lag between changes in the continuum and the emission-line response. In turn, the time lag $\tau_{cent}$ measures mean travel time from the SMBH to the line-emitting region, where $R$ $\sim$ $c \tau_{cent}$ (Koratkar \& Gaskell \citeyear{Koratkar1991}). The method works well for the broad-line region as the region is small and close enough to the photoionizing source to exhibit emission-line variability in response to changes in the continuum. Successful calibrations of the uncertain parameters in the photoionization method based on RM data have permitted for more accurate estimates (Wandel et al. \citeyear{Wandel1999}). This has culminated in several observed relationships such as the $R - L$ relation (Kaspi et al. \citeyear{Kaspi1997}) and the $M-\sigma$-relation (Ferrarese \& Merritt \citeyear{Ferrarese2000}), where scatter around these basic relations can reveal much about the underlying physics and the coevolution of SMBHs and hosts.

We come to the final question: since CoCoA can successfully separate the broad-line 
region from the narrow-line region in Type-1 AGN using line fluxes alone, could it be that the $|\rho'|$ of a certain line is directly measuring the relative emissivity-weighted distance of formation?

\begin{figure*}
 \centering
   \includegraphics[scale=.8]{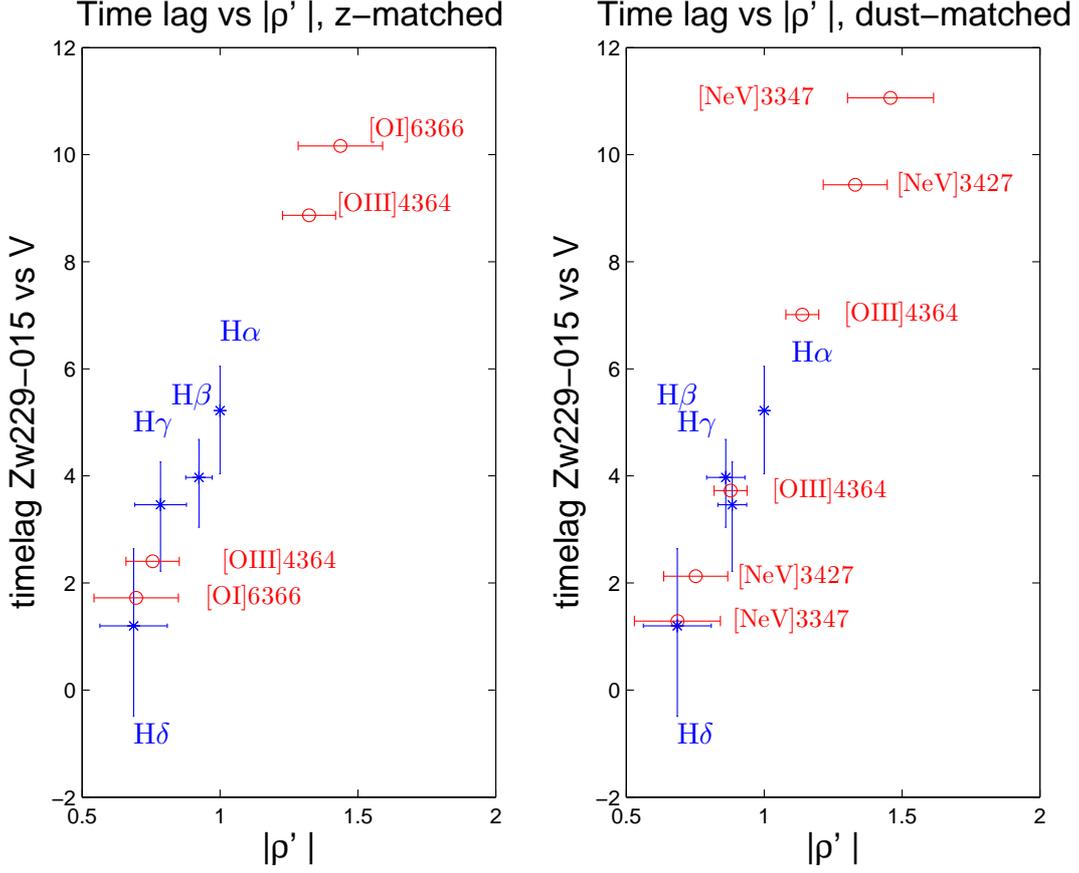}
     \caption{ CoCoA and time lags in Zw229-015. The CoCoA strengths versus the time lags for the Seyfert-1 galaxy Zw229-015 from Barth et al. (2011) are plotted. From the time lags we know the slowest line is H$\alpha$. We use it as our reference point and plot time lags of H$\alpha$, H$\beta$, H$\gamma$ and H$\delta$ versus the $\rho'$(H$\alpha$) for the same lines. The left panel shows the results for redshift-matched Seyfert-1 galaxies (148 objects), the right panel shows the results for dust-matched Seyfert-1 galaxies (87 objects). Blue star symbols show the measured results, their errors representing standard deviations of the mean values. Assuming a linear trend, we also predict time lags for other lines with no time lag information but large $|\rho'$|, shown with red circles. As $|\rho'|$ only is relative to H$\alpha$, the first option is $|\rho'|$ $\times$ $\tau_{cent}$(H$\alpha$) and the second (1/$|\rho'|$)$\times$ $\tau_{cent}$(H$\alpha$). In the left panel, we predict the double-option time lags for [\ion{O}{i}]6366 and [\ion{O}{iii}]4364. In the right panel, we predict the double options for expected time lags for [\ion{O}{iii}]4364 and the two [\ion{Ne}{v}]3347, 3427 lines. For the left panel, the red circles show the predicted double possibilities for time lags for [\ion{O}{i}]6366 and [\ion{O}{iii}]4364. For the right panel, the red circles show the predicted double possibilities for [\ion{O}{iii}]4364, [\ion{Ne}{v}]3347 and [\ion{Ne}{v}]3427. Starting from $|\rho'|$=1 and outwards, note that the double options are always plotted $outwards$ having the same size on their errorbar.}
               \label{TimelagZwicky}
     \end{figure*}

We use the full sample of 148 spiral Seyfert-1s and run CoCoA for H$\alpha$ without any lower limits $|\rho'|$.  We plot the $|\rho'|$ against timelags for the Balmer lines in the Seyfert-1 galaxy Zw229-015 (Barth et 
al. \citeyear{Barth2011}). It is the only AGN in the AGN Black Hole Mass Database (Bentz et al. \citeyear{Bentz2015}) with all four Balmer lines reverberation-mapped. As shown in Fig. \ref{TimelagZwicky} with blue symbols, this gives an apparent relation. Even if our plot only uses four single lines, it suggests that $\rho'$ might be linearly correlated with the emission-weighted distances in AGN. This is interesting as the same technique can then be applied in a similar fashion to extract emission-weighted distances for the narrow-line regions from single-epoch spectra alone, for many galaxies. We repeat the procedure for the dust-matched Seyfert-1s.

\begin{figure*}
 \centering
   \includegraphics[scale=.4]{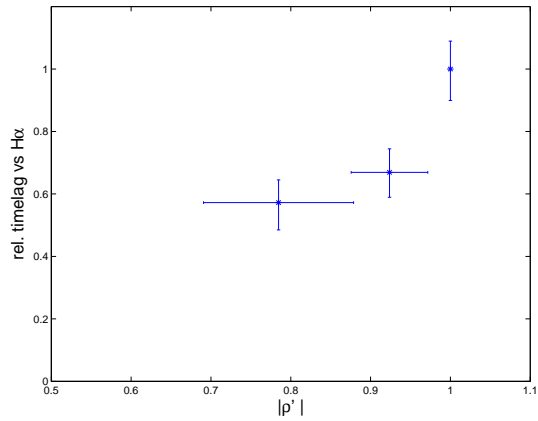}
     \caption{ CoCoA and relative time lags in 13 Seyfert-1 galaxies. The CoCoA strengths versus the relative time lags for 13 Seyfert galaxies from the AGN Black Hole Mass Database. From the time lags we know the slowest line is H$\alpha$. We use it as our reference point and plot relative time lags of H$\alpha$, H$\beta$ and H$\gamma$ versus the $\rho'$(H$\alpha$) for the same lines. All galaxies in the AGN Black Hole Mass Database with time lags for H$\alpha$, H$\beta$ and H$\gamma$ are used, with the exception for Mrk142 as this object has two different sets of time-lag measurements for H$\beta$ with strong disagreement in between. The plotted errors are standard deviations of the mean.}
               \label{TimelagDatabase}
     \end{figure*}

We also use the AGN Black Hole Mass Database to extract relative time-lag data for 13 galaxies with only three line measurements. The results can be seen in Figure \ref{TimelagDatabase} and supports the indicative trend shown in Figure \ref{TimelagZwicky}. The apparent trend in both figures supports the observed ionization stratification of the broad-line region, where the highest ionization emission lines respond the quickest to changes in the continuum (Clavel et al. \citeyear{Clavel}). More galaxies with many lines measured through reverberation mapping are needed to confirm this apparent relation between CoCoA $|\rho'|$ and time lags.

We note that CoCoA returns large $|\rho'|$ for a few narrow lines suggesting a near-BLR origin, also included in Figure \ref{TimelagZwicky} (in red), assuming they follow the same linear trend. Considering the assumed strong dichotomy of the BLR and the NLR, the suggested near-BLR origin for [\ion{O}{i}]6366, [\ion{Ne}{v}]3347,3427 and [\ion{O}{iii}]4364 is a somewhat surprizing result. Does this mean that CoCoA is producing spurious results? Or could it mean that certain narrow lines are formed already in the broad-line region or in its outskirts? Indeed, there are some studies that question the dichotomy and indicate the existence of a transitional region between 
the BLR and NLR, where the gas is intermediate in density and the only true boundary of the BLR is the dust sublimation radius. Transitional lines as the [\ion{Ne}{v}] lines may form here (Murayama \& Taniguchi \citeyear{Murayama}). An example of a galaxy that shows [\ion{O}{iii}]5007 on a much smaller scale than the NLR scale is NGC 5548 where the [\ion{O}{iii}] emission comes from a compact region not larger than a few parsec (Kraemer et al. \citeyear{Kraemer}, Peterson et al. \citeyear{Peterson2013}).

Further support for a transitional line region comes from measuring weak, broad forbidden lines as [\ion{O}{iii}]4364, 5007 and the discovery of broad [\ion{O}{iii}] 
wings in Seyfert galaxies (van Groningen \& de Bruyn \citeyear{Ernst}). More recently, also Balmaverde et al. (\citeyear{Balmaverde}) report results from HST observations showing that [\ion{O}{iii}]4364 can be formed in the inner wall of the obscuring torus. They also find that the velocity dispersion of 
[\ion{O}{i}] is somewhat higher than the velocity dispersion of [\ion{S}{ii}] (740 km/s vs 495 km/s). But the intermediate
line region might be something more complex than a continous transitional gas region, and the extreme broadening of [\ion{O}{iii}]5007
observed in high-redshift quasars suggests strong outflows (Zakamanska et al. \citeyear{Zakamanska}). Given that structures are not 
spherical in AGN, the sight line must also play an important role for observing the transitional line region and for explaining why
the transitional region only has been observed in some objects and not in others.

So does the presence of [\ion{Ne}{v}]3347,3427, [\ion{O}{i}]6366 or [\ion{O}{iii}]4364 in the assumed ``CoCoA-BLR'' 
add to the support of a transitional line region? Great care must be taken to answer this question. The SDSS spectra have 
low resolution and effects from blending and line fitting might bring artifacts (such as false correlations between blended lines) 
into the flux measurements. As discussed by Barth et al. (\citeyear{Barth2016}), such systematic errors in reverberation mapping can result in 
reports of variable forbidden lines. In this work, we refrain from drawing firm conclusions and merely 
illustrate what CoCoA potentially might reveal once the systematic errors are properly accounted for.

Assuming that all line emission arises in one well-behaving cloud and that systematic errors are not present in the flux measurements, 
it can be tempting to predict the time lags of the forbidden lines from the possibly linear behaviour in Figure \ref{TimelagZwicky}.
From $|\rho'|$ we only get the relative time difference with respect to H$\alpha$, but no information on whether the time-lag is shorter or longer, and we 
therefore get two possibilities for the time-lag. The different options for [\ion{Ne}{v}]3347,3427, [\ion{O}{i}]6366 or [\ion{O}{iii}]4364 are plotted with red circles 
in Figure \ref{TimelagZwicky}. Doing this for [\ion{O}{i}]6366 that appears in the redshift-matched samples with $|\rho'|$=0.6959 relative to H$\alpha$, we get that [\ion{O}{i}]6366 should vary at time-scales roughly 0.7 $\times$ $\tau_{cent}$(H$\alpha$) or (1/0.7)$\times$ $\tau_{cent}$(H$\alpha$).
This line is weak and difficult to observe and in the paper of Barth et al. (2011) no particular attention was given to this line, so no time lag information can be found. Still, the support for the linear relation between time lags and $|\rho'|$ is scarce and even if true, it 
may not hold for any lines arising in an outflowing wind.

\section{Conclusion}

We propose a new method, called CoCoA, for finding which lines originate in the same emission-line region based on large samples of galaxies. In theory, not only emission-line fluxes can be used but any observed fluxes e.g. radio, x-ray, IR, continuum and/or emission-line fluxes.

We apply CoCoA to samples of spiral-host Seyfert-1s and Seyfert-2s matched by different criteria
and compare their estimated narrow-line regions. We compare also the NLRs in spiral-host, elliptical-host
and radio-loud Type-2 AGN. Finally, we explore if the method can also yield emission-weighted distances $R$ for different BLR lines.

We conclude:

\begin{enumerate}

\item Results suggest that the same NLR line composition in dust-matched Seyfert-1s and 2s cannot 
be rejected, as seen in the samples matched by reddening H$\alpha$/H$\beta$.

\item The results propose the NLR line composition in Type-2 AGN depends on $both$ host morphology and radio loudness. No difference in NLR electron density based on the sulphur line ratios between spiral and elliptical host Type-2s is found.

\item While CoCoA is suitable to say if a line exists in a sample, it is not suitable for excluding the same possibility. Thus, the method in its present state is not robust enough to do conclusive comparisons of line composition between samples. Meanwhile, the method can be used to confirm the presence of a particular line in an emission-line region, e.g. when the goal is to select the most robust diagnostic tools.

\item There may be a linear relation between the time lags $\tau_{cent}$ from reverberation mapping and CoCoA $|\rho'|$ values. At the moment, the evidence in favour of this relation is sparse. To confirm this possibility, emission-line fluxes and corresponding {expected time lags} for a synthetic set of AGN spectra can be calculated with a photoionization code e.g. \textsc{Cloudy} and analyzed with CoCoA (left for the future). Alternatively, one can test if the predicted time lags from CoCoA agree with those measured in reverberation mapping.

\end{enumerate}

CoCoA's weaknesses may lie in the chosen lower limits, $c_{lim}$ and $c'_{lim}$, and methods of sample selection. 
Selection effects can influence and skew the underlying flux distributions and the estimates of $|\rho|$ and $|\rho'|$ with accompanying errors.
Here the importance of the selected lower limits will act out. More suitable sampling and error calculations can increase
the accuracy of the method that currently suffers from large uncertainties in $|\rho'|$. Assuming the early results
from CoCoA's test cases prove to be correct, an extensive effort to improve CoCoA might be worthwhile -- and 
especially so if it moreover yields relative emission-weighted distances $R$ of various lines as a bonus.

The basic principle of CoCoA is best tested on important scientific cases such as the origin of Fe II lines (Baldwin et al. \citeyear{Baldwin2004})
or the origin of soft X-ray emission in large samples ($n$ $>$ 100) of galaxies (Bianchi et al. \citeyear{Bianchi2005}). If the soft 
X-ray emission really forms in the narrow-line region as proposed, we should observe this as $|\rho'| >$ 0.7 for the 
soft X-ray flux (using [\ion{O}{iii}]5007 as a reference point); likewise for the hypothesis
that Fe II is formed in the broad-line region.

CoCoA allows for statistically ``dissecting'' the AGN components using samples of spectra. With CoCoA one can 
either just find the least contaminated lines for AGN diagnostics, or see if an interesting line forms in 
the same emission-line region as other lines. Resolving the unresolvable through statistics, it provides
a data-driven complement to both theoretical modeling and expensive high-resolution observational studies 
in AGN astronomy. Applying CoCoA on high-resolution ALMA or HST data of Seyfert-1 nuclei may allow us to 
probe the emission-line structure of these nuclei deeper than ever done before.

\begin{acknowledgements}
The authors thank the helpful referee for constructive and important comments. B.V. also thanks E. van Groningen for helpful comments, Brad Peterson for showing the AGN Black Hole Mass Database and A. Magnard, F. Poulenc, G. von Rivia and J. Pianist for inspiring discussions. She also acknowledges A. Barth, H. Netzer and M. Elitzur for help with the intermediate line region.

B.V. was funded and supported by the Center of Interdisciplinary Mathematics (Uppsala Universitet) and Erik and M\"arta Holmbergs donation from the Kungliga Fysiografiska S\"allskapet.

This research also heavily relies on the Sloan Digital Sky Survey (SDSS). Funding for SDSS-II has been provided by the Alfred P. Sloan Foundation, the Participating Institutions, the National Science Foundation, the U.S. Department of Energy, the Japanese Monbukagakusho, and the Max Planck Society.
\end{acknowledgements}

%

\begin{thebibliography}{}
\bibitem[Abazajian et al.(2009)]{Abazajian2009}
Abazajian, K., 2009, The Astrophysical Journal Supplement Series, 182, 543

\bibitem[Antonucci(1993)]{Antonucci1993}
Antonucci, R., 1993, Annual review of astronomy and astrophysics, 31, 473

\bibitem[Baldwin et al.(2005))]{Baldwin2004}
Baldwin, J.E., Ferland G.J., Hamann F. \& Korista K.T., 2004, AGN Physics with the Sloan Digital Sky Survey ASP Conference Series, ed. G.T. Richards \& P.B. Hall, 311

\bibitem[Balmaverde et al.(2015)]{Balmaverde}
Balmaverde, B., Capetti, A., Moisio, D., Baldi R.D. \& and Marconi, A., 2015, Astronomy \& Astrophysics, 586, A48

\bibitem[Barth et al.(2011))]{Barth2011}
Barth, A.J. et al., 2011 The Astrophysical Journal, 732, 121

\bibitem[Barth \& Bentz(2016))]{Barth2016}
Barth, A.J. \& Bentz, M.C., 2016, Monthly Notices of the Royal Astronomical Society Letters, 458, L109

\bibitem[Baskin \& Laor(2005)]{Baskin2005}
Baskin, A. \& Laor A., 2005, Monthly Notices of the Royal Astronomical Society, 358, 1043

\bibitem[Baum et al.(1995)]{Baum1995}	
Baum, S., Zirbel, E.L., O'Dea, C.P., 1995, The Astrophysical Journal, 451, 88

\bibitem[Bentz \& Katz(2015))]{Bentz2015}
Bentz, M.C., Katz, S., 2015, Publications of the Astronomical Society of the Pacific, 127, 67

\bibitem[Bianchi et al.(2005))]{Bianchi2005}
Bianchi, S. , Guainazzi, M. \& Chiaberge M., 2005, Astronomy \& Astrophysics, 448, 499

\bibitem[Brotherton et al.(1994))]{Brotherton}
Brotherton, M.S., Wills, B.J., Francis, P.J. \& Steidel, C. C., 1994, The Astrophysical Journal, 430, 495

\bibitem[Clavel et al.(1991))]{Clavel}
Clavel J., Reichert G.A., Alloin D., et al., 1991, The Astrophysical Journal, 366, 64

\bibitem[Dopita \& Sutherland(1995)]{Dopita1995}
Dopita M.A. \& Sutherland R.S., 1995, Astrophysical Journal, 455, 468

\bibitem[Dultzin-Hacyan et al.(1999)]{Dultzin}
Dultzin-Hacyan et al., 1999, The Astrophysical Journal Letters, 513, L111

\bibitem[Elitzur(2012)]{Elitzur2012}
Elitzur, M., 2012, Astrophysical Journal, 747, p. L33

\bibitem[Ferrarese \& Merritt(2000)]{Ferrarese2000}
Ferrarese, Laura \& Merritt, David, 2000, The Astrophysical Journal, 539, L9

\bibitem[Gaskell(1988)]{Gaskell1988}
Gaskell C.M., 1988, Astrophysical Journal, 325, 114

\bibitem[Jiang et al.(2016)]{Jiang2016}
Jiang N., Wang, H., Mo, H., Dong, X., Wang, T. \& Zhou, H., submitted, arXiv:1602.08825v1

\bibitem[Kaspi(1997)]{Kaspi1997}
Kaspi, S., 1997, in ``Emission Lines in Active Galaxies: New Methods and Techniques'', ed. B. M. Peterson, F.-Z. Cheng,\& A. S. Wilson (San Francisco: ASP), 159

\bibitem[Kauffmann et al.(2003)]{Kauffmann2003}
Kauffmann G. et al. 2003, \mnras, 346, 1055

\bibitem[Kewley et al.(2006)]{Kewley}
Kewley L., Groves, B., Kauffmann G. and Heckman T., 2006, Monthly Notices of the Royal Astronomical Society, 372, 961

\bibitem[Koratkar \& Gaskell(1991)]{Koratkar1991}
Koratkar, A. P., \& Gaskell, C. M., 1991, The Astrophysical Journal Supplement Series, 75, 719

\bibitem[Koulouridis et al.(2013)]{Koulouridis2013}
Koulouridis, E., Plionis, M., Chavushyan, V., Dultzin, D., Krongold, Y., Georgantopoulos, I. \& León-Tavares, J., 2013, 552, 135

\bibitem[Kraemer et al.(1998)]{Kraemer}
Kraemer, S.B., Crenshaw, D.M., Filippenko, A.V., \& Peterson, B.M. 1998, ApJ, 499, 719

\bibitem[Lintott et al.(2008)]{Lintott}
Lintott et al., 2008, Monthly Notices of the Royal Astronomical Society, 389, 1179

\bibitem[Lintott et al.(2010)]{Lintott2010}
Lintott et al., 2011, Monthly Notices of the Royal Astronomical Society, 410, 166

\bibitem[Maiolino et al.(1995)]{Maiolino1995}
Maiolino, R.; Ruiz, M.; Rieke, G. H.; Keller, L. D., 1995, Astrophysical Journal, 446, 561

\bibitem[Murayama \& Taniguchi(1998)]{Murayama}
Murayama T. \& Taniguchi Y., 1998, The Astrophysical Journal Letters, 497, L9

\bibitem[Netzer(1990)]{Netzer1990}
Netzer, H., 1990, in ``Active Galactic Nuclei, Saas-Fee Advanced Course 20'', ed. T. J.-L. Courvoisier \& M. Major (Berlin: Springer), 57

\bibitem[Osterbrock(1988)]{Osterbrock}
Osterbrock, D.E., ``Astrophysics of Gaseous Nebulae and Active Galactic Nuclei'', University Science Books, ISBN 0-935702-22-9

\bibitem[Peterson(1993)]{Peterson1993}
Peterson, B.M., 1993, Publications of the Astronomical Society of the Pacific, 105, 247

\bibitem[Peterson et al.(2013)]{Peterson2013}
Peterson, B.M., Denney, K.D., De Rosa, G., Grier, C.J., Pogge, R.W., Bentz, M.C., Kochanek, C.S., Vestergaard, M., Kilerci-Eser, E., Dalla Bonta, E. \& Ciroi, S., 2013, The Astrophysical Journal, 779, 109

\bibitem[Ramos-Almeida et al.(2011)]{Cristina}
Ramos Almeida, C.; Levenson, N. A.; Alonso-Herrero, A.; Asensio Ramos, A.; Rodríguez Espinosa, J. M.; Perez García, A. M.; Packham, C.; Mason, R.; Radomski, J. T.; Diaz-Santos, T, 2011, Astrophysical Journal, 731, 92

\bibitem[Ricci et al.(2011)]{Ricci}
Ricci, C.; Walter, R.; Courvoisier, T. J.-L.; Paltani, S., 2011, Astronomy \& Astrophysics, 532, A102

\bibitem[Tristram et al.(2014)]{Tristram2014}
Tristram, K.R.W., Burtscher, L., Jaffe, W., Meisenheimer, K., Honig, S.F., Kishimoto, M., Schartmann, M., Weigelt, G., 2014, Astronomy \& Astrophysics, 563, A82

\bibitem[van Groningen \& de Bruyn (1988)]{Ernst}
van Groningen, E. \& de Bruyn, A.G., 1988, Astronomy \& Astrophysics, 211, 293

\bibitem[Villarroel \& Korn(2014)]{VillarroelKorn2014}
Villarroel B. \& Korn A.J., 2014, Nature Physics, 10, 417

\bibitem[Villarroel \& Nyholm et al.(submitted)]{VillarroelNyholm2015}
Villarroel B., Nyholm A., Karlsson T., Comer\'on, S., Korn A., Sollerman J. \& Zackrisson E., 2015, submitted

\bibitem[Wandel et al.(1999)]{Wandel1999}
Wandel, A., Peterson B.M. \& Malkan M.A., 1999, The Astrophysical Journal, 526, 579

\bibitem[Willett et al.(2013)]{Willett}
Willett, K.W., Lintott, C.J., Bamford, S.P., Masters, K.L., Simmons, B.D., Casteels, K.R.V., Edmondson, E.M., Fortson, L.F., Kaviraj, S., Keel, W.C., Melvin, T., Nichol, R.C., Raddick, M.J., Schawinski, K., Simpson, R.J. Skibba, R.A., Smith, A.M. \& Thomas, D., 2013, Monthly Notices of the Royal Astronomical Society, 435, 2835

\bibitem[York et al.(2000)]{York2000}
York, D.G. et al., 2000, Astronomical Journal, 120, 1579

\bibitem[Zakamanska et al.(2015)]{Zakamanska}
Zakamska, N.L, Hamann, F., Paris, I., Brandt, W.N., Greene, J.E., Strauss, M.A., Villforth, C., Wylezalek, D., Alexandroff, R.M. \& Ross, N.P., Monthly Notices of the Royal Astronomical Society, 2016, 459, 3144

\end{thebibliography}
%

\end{document}